\PassOptionsToPackage{a4paper,margin=2.2cm}{geometry}

\documentclass[pdflatex,sn-nature]{sn-jnl}

\usepackage{graphicx}%
\DeclareGraphicsExtensions{.pdf,.png,.jpg,.jpeg}
\usepackage{multirow}%
\usepackage{amsmath,amssymb,amsfonts}%
\usepackage{amsthm}%
\usepackage{mathrsfs}%
\usepackage[title]{appendix}%
\usepackage{xcolor}%
\usepackage{textcomp}%
\usepackage{manyfoot}%
\usepackage{booktabs}%
\usepackage{algorithm}%
\usepackage{algorithmicx}%
\usepackage{algpseudocode}%
\usepackage{listings}%
\usepackage{float}%
\usepackage{upgreek}

\usepackage{url}
\makeatletter
\def\@bibURLprefix{}
\makeatother

\theoremstyle{thmstyleone}%
\theoremstyle{thmstyletwo}%

\theoremstyle{thmstylethree}%

\begin{document}

\title{Controllable Superconductivity in Suspended van der Waals Materials}


\author[1]{\fnm{Ruihuan} \sur{Fang}}
\author[2]{\fnm{Cuiju} \sur{Yu}}
\author[1]{\fnm{Youqiang} \sur{Huang}}
\author[2]{\fnm{Tosson} \sur{Elalaily}}
\author[2]{\fnm{Yuvraj} \sur{Chaudhry}}
\author[1]{\fnm{Yaoqiang} \sur{Zhou}}
\author[4]{\fnm{Andres} \sur{Castellanos-Gomez}}
\author[3]{\fnm{Sanshui} \sur{Xiao}}
\author[5]{\fnm{Jiwon} \sur{Park}}
\author[5]{\fnm{Hyunyong} \sur{Choi}}
\author[1]{\fnm{Fida} \sur{Ali}}
\author*[3]{\fnm{Hanlin} \sur{Fang}}\email{hanfang@dtu.dk}
\author*[2]{\fnm{Jose} \sur{Lado}}\email{jose.lado@aalto.fi}
\author*[2]{\fnm{Pertti} \sur{Hakonen}}\email{pertti.hakonen@aalto.fi}
\author*[1]{\fnm{Zhipei} \sur{Sun}}\email{zhipei.sun@aalto.fi}

\affil[1]{\orgdiv{QTF Centre of Execellence, Department of Electronics and Nanoengineering}, \orgname{Aalto University}, \orgaddress{\city{Espoo}, \postcode{02150}, \country{Finland}}}

\affil[2]{\orgdiv{QTF Centre of Execellence, Department of Applied Physics}, \orgname{Aalto University}, \orgaddress{\city{Espoo}, \postcode{02150}, \country{Finland}}}

\affil[3]{\orgdiv{Department of Electrical and Photonics Engineering}, \orgname{Technical University of Denmark}, \orgaddress{\city{Copenhagen}, \postcode{610101},\country{Denmark}}}

\affil[4]{\orgdiv{2D Foundry Research Group}, \orgname{Instituto de Ciencia de Materiales de Madrid (ICMM-CSIC)}, \orgaddress{\postcode{28049}, \city{Madrid}, \country{Spain}}}

\affil[5]{\orgdiv{Department of Physics}, \orgname{Seoul National University}, \orgaddress{\city{Seoul}, \postcode{08826}, \country{Korea}}}


\abstract{
    Tunable superconductors provide a versatile platform for advancing next-generation quantum technologies. Here, we demonstrate controllable superconductivity in suspended NbSe$_2$ thin layers, achieved through local strain and thermal modulation of the superconducting state. Our results show that suspended NbSe$_2$ structures enable strain modulation of the critical temperature by up to $\sim$0.92~K ($\sim$12.5\% of the critical temperature) and allow the realization of gate-tunable superconducting critical currents. We further demonstrate configurable hysteretic transport characteristics exhibiting multistability and negative differential resistance, providing easily reconfigurable, spatially-dependent superconducting states.
    These phenomena are well explained by calculations of electron–phonon coupling using density functional theory, together with time-dependent Ginzburg–Landau dynamics coupled to the thermal diffusion equation. Our work provides profound insight into strain and thermal modulation of van der Waals superconductors and opens new opportunities for tunable on-chip superconductor devices, integrated superconducting circuits, and quantum simulators.
}





\maketitle

\section*{Introduction}\label{sec1}

\quad\quad Controllable superconductors will be central to future quantum technologies~\cite{Gallagher2014}. Van der Waals two dimensional (2D) superconductors provide a versatile platform for tunable superconductivity, offering multiple tuning knobs, such as gate doping~\cite{Ruf2024a, Ruf2024, Elalaily2021, Elalaily2024, Chen2019, Uri2023, Lu2015}, layer thickness~\cite{Uchihashi2017,Dvir2018}, pressure \cite{Rahman2022}, and strain~\cite{Wieteska2019}. These methods can modulate intrinsic superconductor properties, such as the critical temperature ($T_c$) and critical current ($I_c$). Among these methods, strain engineering is particularly attractive because it directly modifies lattice symmetry and electronic bands, enabling continuous and reversible control of superconductivity~\cite{Chen2020, Wieteska2019, Henriquez-Guerra2024}. Most existing strategies to tune strain in two-dimensional superconductors rely on the substrate—for instance, by exploiting piezoelectric effects~\cite{Wieteska2019}, thermal expansion or contraction~\cite{Henriquez-Guerra2024}, or lattice mismatch in epitaxial growth~\cite{Chen2020}. These methods, however, intrinsically couple strain to other parameters such as temperature or crystal orientation, and typically provide limited post-fabrication tunability. In contrast, electrostatic modulation in freely suspended 2D superconductors enables reversible and continuous strain control using a simple gate voltage~\cite{lemmeNanoelectromechanicalSensorsBased2020}, offering a dynamic and purely electrical means to tune the mechanical state of the material~\cite{houTuningInstabilitySuspended2024}, and achieving strain levels that surpass those attainable with piezoelectric substrates~\cite{Rebollo_2021, naumisElectronicOpticalProperties2017}. Just as importantly, suspended structures strongly suppress heat transfer into the substrate by eliminating substrate interference, thereby enhancing the thermal response of superconductors compared to the supported geometries, where substrate-mediated thermal transport remains dominant and hinders large thermal tunability~\cite{Jing2018}.
Furthermore, since the suspended structure requires only a simple etching process, it enables additional tunability through suspended patterns in 2D superconductors, which can be precisely engineered using advanced micro- and nanofabrication techniques. This makes suspended 2D superconducting devices an excellent platform for exploring coupled thermal–superconducting phenomena~\cite{Du2009, Feldman2009, Locatelli2010, Wu2012, Jin2015, Gurevich1987, Reiss2014, Li2022}.
However, controllable superconductivity in suspended 2D superconductors has remained largely unexplored.

Here, we demonstrate the controllability of the superconducting state in suspended NbSe$_2$ thin layers through engineered strain and controlled thermal response. Locally tunable strain–driven controllable superconductivity is realized through a gate-induced deformation mechanism in the suspended structure, allowing dynamic and spatially selective control of $T_c$ and, correspondingly, $I_c$. In parallel, strong and spatially resolved control of the thermal response in superconducting transport is achieved through the enhanced thermal response of the suspended structure, allowing the realization of the nonlinear multi-stage hysteresis and negative differential resistance (NDR) in the transport curves. These observations are supported by first-principles calculations linking strain to reduced electron–phonon coupling and superconducting gap, and by time-dependent Ginzburg–Landau (TDGL) simulations coupled to thermal diffusion, which capture current-driven switching and delayed recovery. Our results demonstrate that strain (for threshold setting) and thermal feedback (for state control) can be harnessed in suspended 2D superconductors, opening opportunities for on-chip tunable superconducting elements, integrated circuits, and quantum simulation platforms.

\section*{Results}\label{sec2}
\subsection*{The concept of controllable superconductivity with suspended NbSe$_2$}\label{secconcept}
\begin{figure}[htb]
    \centering
    \includegraphics[width=1.0\linewidth]{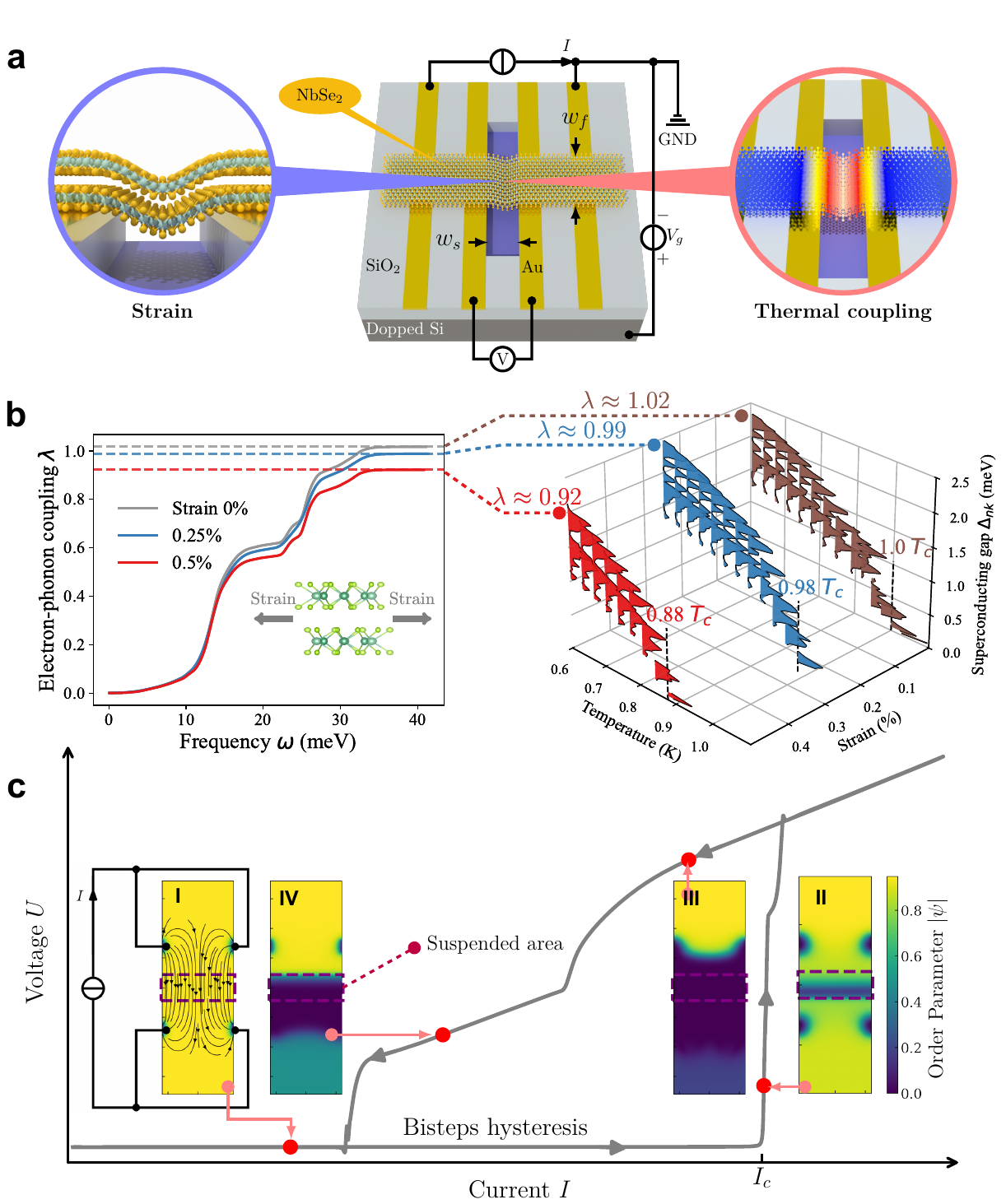}
    \caption{\textbf{Controllable suspended NbSe$_2$ superconductor concept with strain and thermal modulation}. (a) The schematic of a suspended controllable NbSe$_2$ superconductor. The NbSe$_2$ flake with flake width $w_f$ is transferred onto a SiO$_2$ substrate containing prepatterned 4 electrodes and an opening (indicated by purple color) that defines the suspended area with suspended width $w_s$.
    The suspended area can be modulated by tensile strain when applying a gate voltage $V_g$ for controllable superconductivity (Left inset).
    When applying the bias current $I$, the heat generated by current dissipation will interact with the superconducting state (Right inset).
    (b) Strain-tunable superconductivity. Left panel indicates that the calculated electron-phonon coupling strength as a function of frequency in strain of 0\%, 0.25\% and 0.5\% with the corresponding stable electron-phonon coupling strengths of $\sim$ 1.02, 0.99, and 0.92, respectively.
     The right panel shows that the calculated superconducting gap $\Delta_{nk}(w = 0)$ as a function of temperature under strain of 0\%, 0.25\%, and 0.5\%. The corresponding critical temperatures are $T_c$, 0.98 $T_c$, and 0.88 $T_c$, respectively, where $T_c$ is the critical temperature of the unstrained NbSe$_2$.
    (c) Controllable superconductor with thermal coupling. The simulated $I$-$V$ curve and the order paratmeter heatmap of the suspended NbSe$_2$ with a suspended width $w_s=$ 8.8 $\xi$ marked as purple dashed boxes.
    }
    \label{fig:schematic}
\end{figure}

\quad\quad The schematic of our controllable superconductivity concept with suspended NbSe$_2$ is shown in Fig.~\ref{fig:schematic}a.
Four electrodes (shown in yellow) form a four-terminal probe — the outer pair injects current, while the inner pair measures voltage.
A rectangular trench with a width of $w_s$ is etched between the inner electrodes, and an NbSe$_2$ thin flake of width $w_f$ is transferred to span the etched trench, making contact with all four electrodes to form a suspended NbSe$_2$ device with a suspended width $w_s$ for four-terminal measurements.
Depending on the etched trench pattern, the suspended structure can provide controllable strain and thermal coupling. When a gate voltage $V_g$ is applied via the doped silicon back-gate, the portion of the NbSe$_2$ flake above the trench (i.e., the suspended part) deforms downward due to electrostatic attraction, thereby inducing strain in the suspended NbSe$_2$ part.

Under an applied current through the outer contact pair, Joule heating arises in NbSe$_2$ and interacts with its superconducting state. Here, in our suspended NbSe$_2$ geometry, heat is retained within the superconductor due to its weak thermal conductivity, enhancing the thermal response of superconductors. This is different from previously reported results for 2D superconductors supported on substrates~\cite{Chen2021,Wu2012}, where heat is typically conducted into the substrate, thereby limiting the thermal–superconductor coupling, i.e., intrinsic thermal modulation of the superconductor.

To investigate how strain affects the superconducting properties, we first examine its effect on the electron–phonon coupling strength within the Bardeen–Cooper–Schrieffer framework via DFT calculation~\cite{Giustino2007,Eliashberg1960}. As shown in Fig.~\ref{fig:schematic}b, the calculations (see Methods for details) yield the electron–phonon coupling coefficient $\lambda(\omega)$ under applied strains of 0\%, 0.25\%, and 0.5\%. The inset in the left panel of Fig.~\ref{fig:schematic}b shows a schematic representation of how strain is applied to the NbSe$_2$ lattice. With increasing strain, $\lambda$ decreases from $\sim$1.02 to 0.99 and 0.92, leading to a reduction in the superconducting gap. In theory~\cite{Eliashberg1960}, the superconducting energy gap $\Delta_{\mathrm{nk}}$ protects the electron–phonon coupling that binds electrons into Cooper pairs, which are essential for superconductivity. Therefore, a larger superconducting energy gap generally corresponds to a higher critical temperature. Consequently, the strain-induced reduction in the superconducting energy gap correspondingly modulates the critical temperature from $T_c$ (without strain) to $\sim$ $0.98T_c$ and $0.88T_c$, respectively, as illustrated in the right panel of Fig.~\ref{fig:schematic}b.  Therefore, strain in suspended NbSe$_2$, induced through electric gate tuning, provides a viable method for dynamically controlling superconductivity.

In addition to the above-mentioned strain-controlled electron–phonon interactions, intrinsic thermal coupling in suspended geometries provides a second tunable parameter that influences superconductivity through  the configured suspended geometry.
The superconductor–thermal coupling is quantitatively captured by coupling the TDGL equation with the heat-diffusion equation (Details in Methods). Heat generation mainly arises from two sources:
On one hand, the formation of the superconductor cooper pairs will produce heat since the cooper pairs is under low energy ground state~\cite{Kramer1978,Hernandez2008};
On the other hand, when the transport current exceeds the critical current, the superconductor becomes a normal metal and produces Joule heat due to the finite resistance.
Since the suspended area prevents thermal dissipation through the substrate, the heat-transfer coefficient $\eta$ (See Eq. (5) of Methods ) of the suspended area is much lower than that of the supported area, leading to a strong thermal response in the suspended region.
In our simulation, we set $\eta=0.05$ for the suspended region and $\eta=10$ for the supported region. For comparison, Ref.~\cite{Jing2018} used $\eta=2\times10^{-3}$ when modeling with a low heat transfer coefficient. Our choice of $\eta=0.05$ is more conservative (i.e., it allows stronger thermal exchange) because in our cryostat, residual helium gas provides some heat transfer to the environment, unlike the high-vacuum conditions assumed in the reference. For the supported region, $\eta=10$ is chosen to effectively suppress the thermal response in that area, ensuring a clear distinction between thermally coupled and thermally decoupled regions within a single unified simulation framework.
As shown in the inset panel I of Fig.~\ref{fig:schematic}c, a bias current is injected at the upper boundary and extracted at the lower boundary of a flake example with dimensions of 24~$\xi$$\times$64~$\xi$, where $\xi$ is the coherence length of the superconductor order parameter~\cite{Bishop-VanHorn2023}. A suspended region, $w_s=$8.8~$\xi$ in width, crosses the flake and is indicated by purple dotted rectangles.
The black arrow streamlines in the inset panel I indicate the current flow. As the flowing current increases and reaches $I_c$, the suspended region undergoes a sharp transition from the superconducting state to the normal metallic state, as shown in the inset panel II of Fig.~\ref{fig:schematic}c.

After the transition to the normal metallic state, the flowing current generates Joule heat due to the ohmic dissipation. Therefore, in the case of the downward current sweep, the generated heat is trapped in the suspended region, which increases temperature, delaying superconducting recovery and producing a hysteresis loop (see panel III of the inset in Fig.~\ref{fig:schematic}c).
The two-stage hysteresis in the $I$–$V$ curve arises from the spatial dynamics of the superconductivity recovery during the current reduction. Initially, as the current is reduced, the voltage decreases slowly, reflecting a gradual and spatially uneven recovery of the superconducting order parameter. This asymmetry is due to slight imperfections in the finite volume simulation mesh, which causes the order parameter to begin recovering preferentially from one side. In analogy to a real experimental situation, where geometric asymmetries in the flake and spatial variations in defect density similarly break the symmetry and trigger preferential recovery from one region.
As current reduced, the recovery progresses goes faster, and the faster recovery of the superconductors will produced heat.
When the superconductor recovery domain wall forms across suspended area, the heat produced by the recovery of superconductor states will be kept due to low heat transfer rate to substrate, and raise the temperature of the suspended area. 
The raise of the temperature effectively halts further moving of superconductor recovery domain wall, trapping the rest of the device in the resistive state(inset panel IV of Fig.~\ref{fig:schematic}c). At this point, the order parameter distribution stabilizes, and the $I$–$V$ curve transitions to a linear slope — indicating constant resistance with no further phase recovery. This sequential behavior defines the two-stage hysteresis: a partial superconducting recovery followed by a steady resistive regime.
Consequently, through thermal coupling, the suspended NbSe$_2$ device can be controlled to realise three different states: a fully superconducting state (Inset Panel I), a spatially resolved mixed state with coexisting normal and superconducting regions (Inset Panel IV), and a fully normal metallic state (Inset Panel III).

In summary, the suspended NbSe$_2$ platform enables electrically controllable superconductivity: gate-induced strain tunes the electron–phonon coupling $\lambda$, the superconducting energy gap $\Delta$, and thus $T_c$, while the suspended geometry enhances thermal response that, under current bias, drives controlled transitions among fully superconducting, spatially mixed (coexisting normal and superconducting), and fully normal states. This affords deterministic, in situ control of $T_c$, $I_c$, and device states for reconfigurable superconducting circuits.

\subsection*{Controllable superconductivity with electrically tunable strain}

\begin{figure}[htb]
    \centering
    \includegraphics[width=1.0\linewidth]{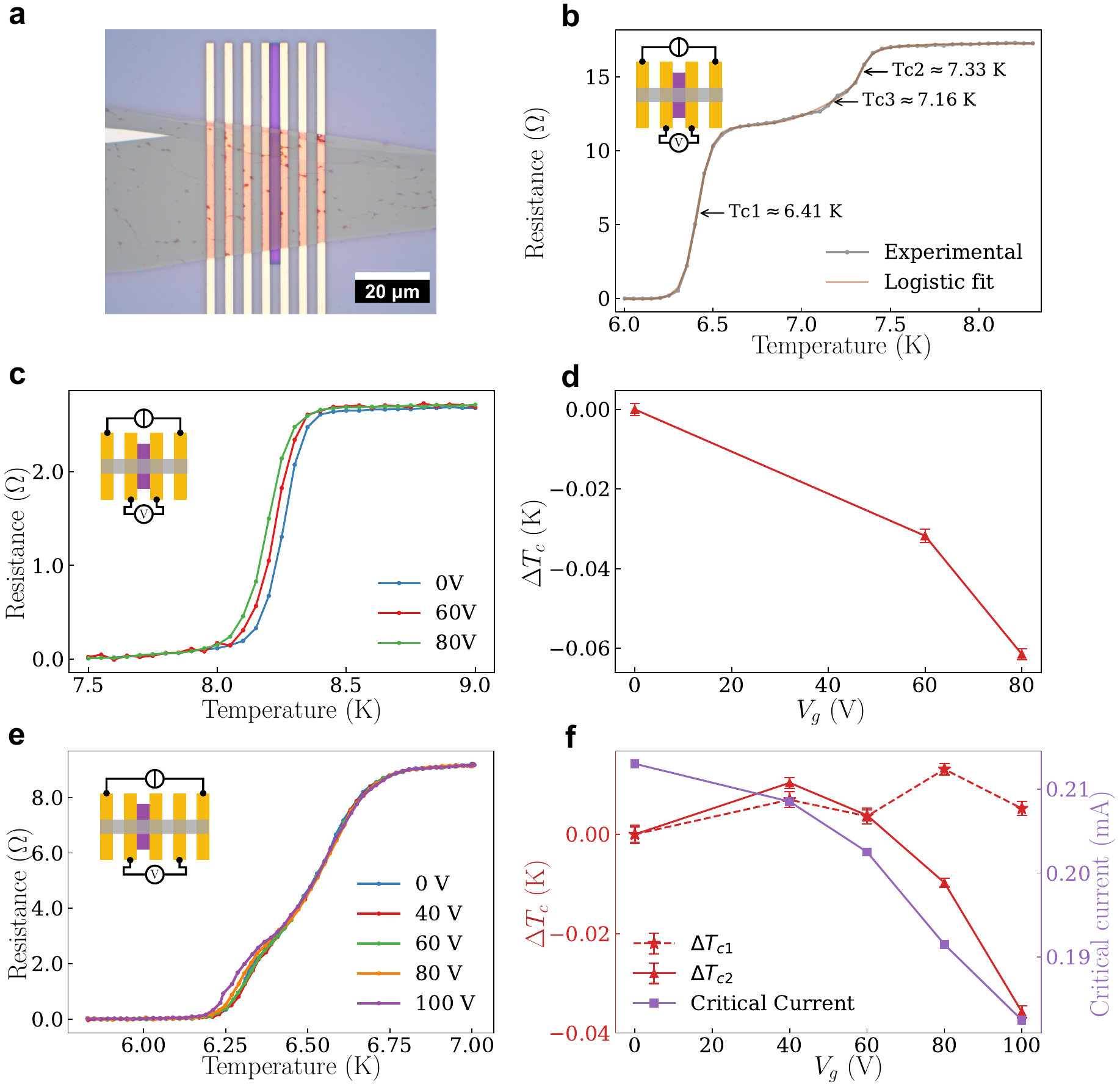}
    \caption{\textbf{Controllable superconductivity with strain.}
    (a) Optical micrograph of a suspended NbSe$_2$ device with multiple electrodes. The purple area outlines a $\sim$600~nm-deep cavity etched into a 1000~nm-thick SiO$_2$ layer; Yellow strips are Au electrodes.
    (b) Resistance–temperature ($R$–$T$) curves of a collapsed and partially suspended NbSe$_2$ structure. Critical temperatures are extracted by fitting a sum of three logistic components. The inset indicates the four-terminal measurement configuration.
    (c) $R$–$T$ curves of a fully suspended NbSe$_2$ device under $V_g=0$, 60, and 80~V. The inset indicates the four-terminal measurement configuration.
    (d) Fitted $T_c$ change $\Delta T_c$ versus $V_g$ for the device in (c).
    (e) $R$–$T$ curves from a geometry where the four-terminal path spans both suspended and SiO$_2$-supported regions, measured under $V_g=0$, 40, 60, 80, and 100~V. The inset indicates the four-terminal measurement configuration.
    (f) Fitted critical temperature changes and critical current versus $V_g$ for the device in (e). The behaviour arises from the combination of electrically tunable suspended superconductivity, where $\Delta T_{c1}$ is modulated by the gate voltage $V_g$, and electrically independent non-suspended superconductors, where $\Delta T_{c2}$ remains nearly constant with respect to $V_g$.
    }
    \label{fig:criticaltemperature}
\end{figure}

Our controllable suspended NbSe$_2$ superconductor device is shown in Fig.~\ref{fig:criticaltemperature}a. A thin flake bridges a trench in SiO$_2$ and, together with multiple Au electrodes (width 2~$\upmu$m, thickness 50~nm), forms a four-terminal layout (details in Methods). In these special channel-based devices, the whole suspended area spanning the NbSe$_2$ flake ensures that the strain is fully applied via the electrical gate. To minimize strain induced by thick electrodes, we employ embedded electrodes so that the level of the electrodes is nearly aligned with the SiO$_2$ top surface (height difference $\sim$5~nm).

To study how strain influences superconductivity, we first characterize a collapsed but partially suspended structure, which exhibits the maximum strain and thus introduces maximal possible change in $T_c$. An optical micrograph of the device is shown in Fig. S2a of the Supplementary Information. The depth of the trench is $\sim$600 nm, according to atomic force microscope measurement and the intended suspended width $w_s\approx 3~\upmu$m. The four-lead transport measurement configuration is shown in the inset of Fig. \ref{fig:criticaltemperature}b.
Owing to multiple inhomogeneous strain domains in this collapsed but partially suspended device, the NbSe$_2$ exhibits multiple critical temperatures. The critical temperature is extracted by fitting the temperature-dependent resistance $R$($T$) using a sigmoid function with three logistic levels (detailed discussion of the strain distribution in Supplementary Section 2), yielding three apparent critical temperatures of $\sim$7.33~K, 7.16~K, and 6.41~K.
The maximum variation in $T_c$ is $\sim$0.92~K (about 12.5\% of the sample's $T_c$), establishing that strain in the suspended architecture can significantly modulate superconductivity. This benchmark sets the upper limit for the controllable tuning range demonstrated in the following gate-controlled devices.

To characterise control of superconductivity  by strain, we measure the temperature-dependent resistance across a fully suspended flake under $V_g=0, 60, 80$~V (Fig.~\ref{fig:criticaltemperature}c).
Upon applying a gate voltage, the suspended NbSe$_2$ region deforms downward due to electrostatic attraction, inducing mechanical strain. As the gate voltage increases, the $R$–$T$ curve shifts progressively to lower temperatures. The sigmoid-fitted $T_c$–$V_g$ relation (Fig.~\ref{fig:criticaltemperature}d) shows that $T_c$ decreases monotonically with increasing $V_g$: with $0\to80$~V, the net change is $\sim$0.06~K (about 1\% of the sample's $T_c$), demonstrating gate-controllable tuning of the critical temperature. Based on the measured gate-induced membrane deformation from optical spectroscopy, we estimate the strain to be $\sim$0.07\% (see Supplementary Section~3 for details). Control measurements confirm that the observed $T_c$ and $I_c$ modulation arises from strain rather than gate-leakage-induced heating, as the critical parameters remain suppressed even when leakage current decreases (Supplementary Section~4).

The coexistence of suspended and substrate-supported segments establishes sharp gradients in both mechanical strain and thermal anchoring, thereby generating distinct local superconducting gaps and critical currents within a single flake. To resolve this controllable superconducting landscape, we measure the four-terminal voltage along a path that threads the tensile suspended span and the clamped supported region (measurement configuration in the inset), as shown in Fig.~\ref{fig:criticaltemperature}e.
The $R$–$T$ curves measured at $V_g=0, 40, 60, 80, 100$~V exhibit two distinct transitions, which show that the two NbSe$_2$ regions are exhibiting distinct $T_c$ values. Two-step sigmoid fits in Fig.~\ref{fig:criticaltemperature}f reveal two distinct $\Delta T_c$ branches, labeled $\Delta T_{c1}$ and $\Delta T_{c2}$. $\Delta T_{c1}$, marked by star symbols, remains nearly constant with increasing $V_g$ and corresponds to the supported region. In contrast, $\Delta T_{c2}$, marked by triangle symbols, decreases with $V_g$ and is associated with the suspended segment, where the superconductivity is tuned by strain that is controllable via the gate voltage. This confirms a spatial separation of superconducting behavior, enabling localized, gate-controlled of $T_c$ within a single device. Concurrently, the critical current $I_c$ decreases with increasing $V_g$, consistent with the reduction in $T_c$. Together, these results demonstrate that gate-induced deformation in a suspended geometry enables dynamic and spatially resolved control of superconducting parameters in NbSe$_2$.

\subsection*{Controllable superconductive transition with thermal modulation}
\begin{figure}[ht]
    \centering
    \includegraphics[width=1.0\linewidth]{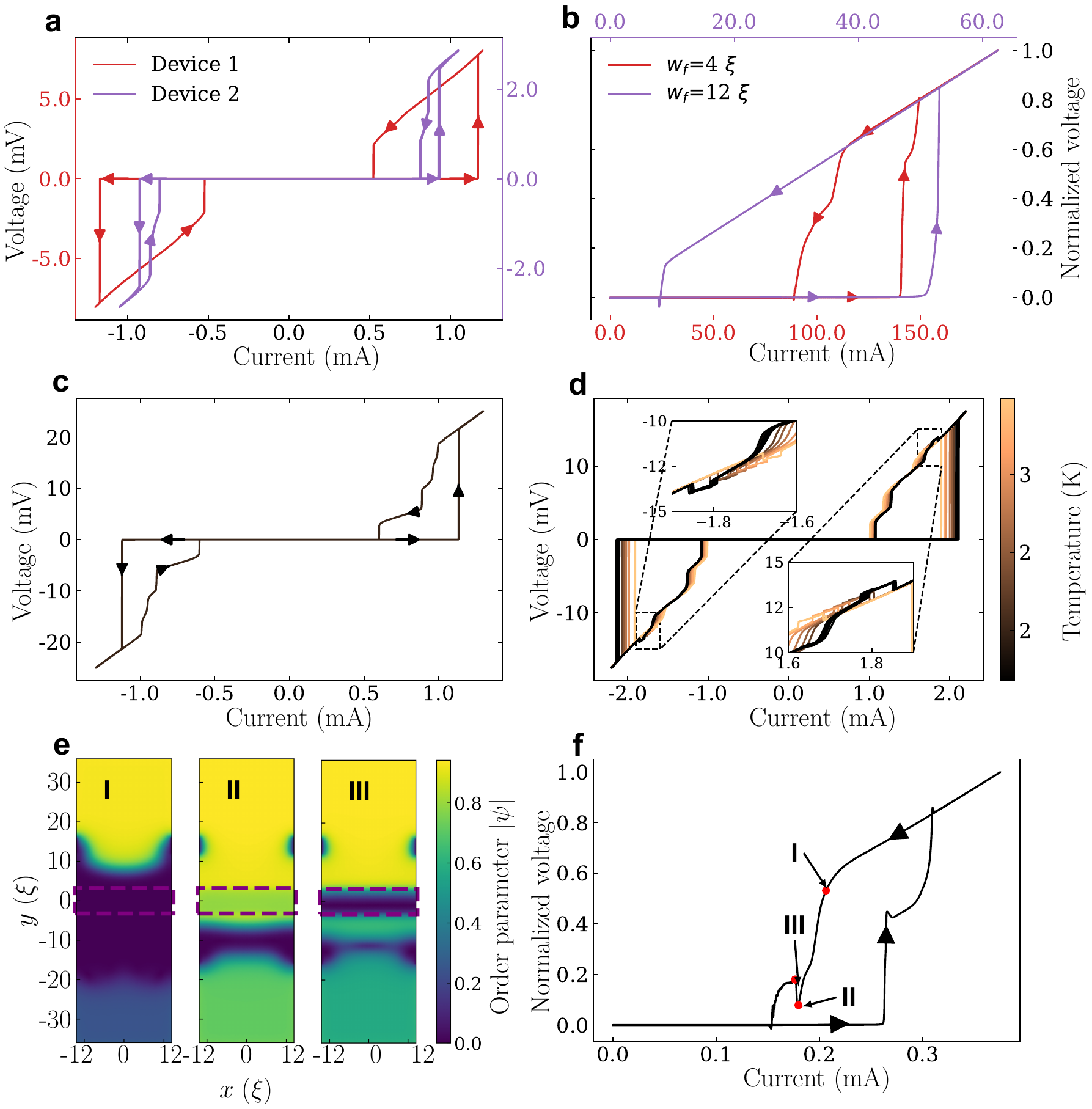}
    \caption{\textbf{Controllable superconductivity with thermal modulation}.
    (a) $I$--$V$ curves of suspended NbSe$_2$ at $3~\mathrm{K}$ showing thermal hysteresis: red line shows single-step hysteresis for Device 1 with flake width $w_f\approx 8~\upmu\mathrm{m}$, suspended width $w_s\approx 3~\upmu\mathrm{m}$, and thickness $\sim 13~\mathrm{nm}$ (Supplementary Fig.~S1a), and blue line shows two-step hysteresis for Device 2 with flake width $w_f \approx 25~\upmu\mathrm{m}$, suspended width $w_s\approx 2~\upmu\mathrm{m}$, and thickness $\sim 11~\mathrm{nm}$ (Supplementary Fig.~S1b).
    (b) TDGL simulations for $64~\xi$-long flakes interacting with a suspended width of $w_s=8.8~\xi$; the resulting $I$--$V$ curves for widths of $w_f=4~\xi$ and $w_f=12~\xi$ are shown in red and blue, respectively.
    (c) $I$--$V$ curves of suspended NbSe$_2$ at $3~\mathrm{K}$ exhibiting multistep hysteresis for Device 3 with flake width $w_f \approx 8~\upmu\mathrm{m}$, suspended width $w_s\approx 3~\upmu\mathrm{m}$, and thickness $\sim 13~\mathrm{nm}$ (Supplementary Fig.~S1c).
    (d) Temperature-dependent $I$--$V$ curves for a device with flake width $w_f \approx 25~\upmu\mathrm{m}$, suspended width $w_s\approx 4~\upmu\mathrm{m}$, and thickness $\sim 11~\mathrm{nm}$ (Supplementary Fig.~S1d); insets highlight regions of negative differential resistance.
    (e) TDGL coupled to the thermal diffusion equation: simulated order parameter $|\psi|$ for suspended with of $w_s=6.4~\xi$, flake width of $w_f=24~\xi$, illustrating states I--III corresponding to distinct $I$--$V$ regimes.
    (f) TDGL+thermal-diffusion simulated $I$--$V$ curve with stages I--III marked, corresponding to the spatial maps in (e).
    }
    \label{fig:iv}
\end{figure}

Using a suspended structure minimizes thermal exchange with the substrate, thereby enhancing the coupling between heat and superconductivity. This stronger thermal–superconducting interaction gives rise to bistable or multistable superconductivity characteristics. To reach an overall picture of these phenomena, we fabricated four different devices with varying suspended widths $w_s$, flake widths $w_f$, and flake thicknesses, and investigated how the suspended device geometry governs the thermal–superconducting coupling strength and its electrical manifestations.

For Device 1, which has the suspended width of $w_s$ $\approx3~\upmu\mathrm{m}$, a flake width of $w_f \approx $ $8~\upmu\mathrm{m}$, and a NbSe$_2$ flake thickness of $\sim13~\mathrm{nm}$ (see micrographs in the Supplementary information). The measured $I$-$V$ curve of the device is shown in Fig.~\ref{fig:iv}a (red solid line). When the current exceeds the critical value ($\sim$ 1.17 mA), the device switches to the normal state and generates Joule heat. On the return sweep, the generated heat under the metallic state prevents the immediate recovery of superconductivity. The current needs to be further reduced to reach a lower retrapping point ($\sim$ 0.5 mA) to restore the superconductivity state, resulting in a clear single-step thermal hysteresis.

Device 2, by contrast, has a smaller suspended width of $w_s\approx$ $2~\upmu\mathrm{m}$, a much wider flake of $w_f\approx$ $25~\upmu\mathrm{m}$, and a thickness of $\sim11~\mathrm{nm}$. It exhibits a two-step hysteresis behavior, as shown by the purple solid line in Fig.~\ref{fig:iv}a. As discussed in subsection~\ref{secconcept}, this two-stage transition reflects spatially inhomogeneous superconducting recovery, where the device partially reenters the superconducting state before the rest due to a thermal boundary.

To further investigate the factors influencing the hysteresis steps, we performed numerical simulations using the TDGL model coupled with the thermal diffusion equation. Although Devices 1 and 2 differ in both flake width and suspended width, we isolate the effect of flake width in our simulations to identify its specific role. In the simulation results (Fig.~\ref{fig:iv}b), we fixed the suspended width at $w_s=8.8~\xi$ and varied the flake width $w_f$. Details about the simulation are shown in the Supplementary Information. As shown in Fig.~\ref{fig:iv}b, narrower flakes with flake width $w_f=4~\xi$ produce single-step hysteresis, while wider flakes with flake width $w_f=12~\xi$ produce two-step transitions.  This result aligns with the above experimental data in Fig.~\ref{fig:iv}a: Narrow Device 1 (with $w_f\approx 8~\upmu\mathrm{m}$) shows a single step, while wide Device 2 (with $w_f\approx 25~\upmu\mathrm{m}$) shows two steps. This is because the recovery of the narrow flake occurs at a current much lower than the critical current $I_c$ compared to the wider flake. At this point, the current is relatively small, and the temperature increase caused by the heat released during the recovery of the superconducting state has little influence on the superconductivity.

In addition to the hysteresis behaviors observed in Devices~1 and~2, Device~3 (a suspended width of $w_s\approx 3~\upmu\mathrm{m}$, a flake width $w_f \approx 30~\upmu\mathrm{m}$, and a thickness of $\sim 5~\mathrm{nm}$; details in Supplementary Fig.~S1c) exhibits a more complex switching pattern with multiple discrete steps (Fig.~\ref{fig:iv}c). The voltage is measured across contacts to form a device that includes a suspended region and a supported part. During the downward current sweep, the I--V curve indicates that different regions of the device return to the superconducting state at different currents, which forms the multistage behaviour arising from the disparate thermal coupling in the suspended and supported regions. Such responses highlight the potential for suspended superconducting structures to exhibit geometry-dependent switching, thus enabling fully controllable hysteresis behaviour.

Beyond inducing spatially inhomogeneous superconducting recovery, thermal coupling can also give rise to NDR~\cite{Misko2007, Ustavcshikov2024}. NDR is a counterintuitive phenomenon in which increasing the applied force results in reduced particle velocity~\cite{Misko2007}, and has been observed in various nonlinear systems, including semiconductors, plasmas, and superconductors~\cite{Pedersen2014, doi:10.1126/sciadv.aba1377}.
In Device~4, which features a suspended width of $w_s\approx 4~\upmu\mathrm{m}$, a flake width $w_f \approx 25~\upmu\mathrm{m}$, and a thickness of $\sim 11~\mathrm{nm}$ (see Supplementary Fig.~S1d), we observe pronounced NDR along the hysteresis path. As highlighted in the zoomed region of Fig.~\ref{fig:iv}d, the voltage first decreases and then unexpectedly increases as the current continues to decrease, forming a distinct NDR segment. We further carry out temperature-dependent measurements by varying the sample temperature, as shown in Fig.~\ref{fig:iv}d. The onset current of NDR shifts to lower values as the cryostat temperature decreases, demonstrating that the NDR behavior is repeatable and can be controlled through temperature.

To understand the NDR mechanism, we performed TDGL simulations coupled with heat transport on a flake geometry with a suspended width $w_s=6.4~\xi$ and a flake width $w_f=24~\xi$. The simulation reproduces the key features: after superconductivity is lost at high current, recovery gradually occurs upon decreasing current (Fig.~\ref{fig:iv}e.\,I–II).
The recovery of the superconductor state in the suspended region produce heat, followed by a reentrant transition to the normal metallic state (blue section) in the central region, causing a voltage rise and forming an NDR segment (Fig.~\ref{fig:iv}e.\,III). This reentrant behavior arises from thermal instability, where the locally elevated temperature in the center prevents superconducting recovery even as the current continues to decrease.
Both experiment and simulation confirm that under extremely weak thermal exchange to the substrate, NDR can emerge even in simple suspended geometries. This demonstrates a practical route to designing nonlinear superconducting elements where hysteresis shape and NDR behavior are controlled through geometry and thermal engineering — offering a foundation for custom nonlinear circuit components and future quantum device platforms.

\section*{Conclusions}\label{sec3}

We have demonstrated that suspended van der Waals superconductors provide a simple and effective platform for controllable superconductivity. In suspended NbSe$_2$ devices, gate-induced deformation generates localized, reversible strain fields that modulate key superconducting parameters such as $T_c$ and $I_c$. At the same time, the suspended geometry reduces thermal coupling to the substrate, enhancing thermal–superconducting interactions and enabling heat-driven control of transport phenomena—including multi-stage hysteresis and NDR.
First-principles calculations using DFT and the Electron-Phonon Wannier (EPW) code reveal that the strain weakens the electron-phonon coupling and suppresses the superconducting gap, explaining the observed gate dependence of $T_c$. These DFT and EPW results align with complementary simulations based on time-dependent Ginzburg–Landau theory coupled to diffusive heat transport, which captures the switching dynamics, delayed recovery, multistage transitions and NDR observed in the experiment.
Together, these results establish suspended 2D materials as a versatile system where superconductivity can be engineered through geometry and gating. Strain defines local switching thresholds, while thermal feedback governs transport pathways, enabling access to engineered $I$–$V$ responses within a single, continuous material. This combination of physical tunability and structural simplicity opens the door to both new regimes of superconducting physics and the development of reconfigurable superconducting circuits, for example for neuromorphic computing.

\section*{Methods}\label{sec4}
\subsection*{Fabrication of suspended NbSe$_2$ devices}
The suspended NbSe$_2$ devices were fabricated using a multi-stage electron beam lithography (EBL) process.
First, for electrode deposition, PMMA A6 was spin-coated on a 1000 nm SiO$_2$/Si substrate at 4000 rpm for 1 minute, followed by EBL exposure. The SiO$_2$ layer was then etched to a depth of 50 nm using reactive ion etching (RIE) with CHF$_3$ and Ar gases. Subsequently, a 5 nm Ti layer was deposited, followed by a 45 nm Au layer, ensuring that the electrodes were nearly flush with the SiO$_2$ surface.
Next, for trench etching, PMMA A9 was spin-coated at 4000 rpm for 1 minute and another EBL exposure was performed. RIE was used to etch 600 nm deep trenches to facilitate the suspended structure.
Finally, NbSe$_2$ was exfoliated onto PDMS and transferred to reside across the trench structure. The transfer was completed by heating to 40$^\circ$C and slowly peeling off the PDMS, allowing the material to settle onto the selected place as a suspended structure.

\subsection*{Transport measurements}

The devices were mounted in the Attocube attoDry 2100, which can achieve temperatures down to 1.6 K.
Transport measurements were performed using a Keithley 6221 precision current source to provide bias current and a Keithley 2182A voltammeter for voltage measurements.
The resistance measurements were conducted in delta mode (current reversal technique to eliminate the influence of thermal voltages) with a delta current amplitude of 1 $\mu$A.
The current is injected through the two outer electrodes, and the voltage is measured between two inner electrodes.
The Keithley 2400 source meter is used to provide the gate voltage to the NbSe$_2$ device.
The critical temperature is obtained by fitting $R(T)$ with a logistic(sigmoid) function,
\begin{equation}
    R(T) = R_0 + \frac{R_1 - R_0}{1 + e^{-(T - T_0)/\omega}},
\end{equation}
where $R_1$ is the normal-state resistance, $R_0$ is the resistance below the superconducting transition temperature (due to other normal regions), $T_0$ is the critical temperature of the fitted section, and $T$ is the bath temperature.

\subsection*{First-principles calculations}

We performed first-principles calculations with QUANTUM ESPRESSO package using relativistic norm-conserving ONCV pseudopotentials~\cite{Giannozzi2009}. The Perdew-Bruke-Ernzerhof exchange-correlation functional of generalized gradient approximation was adopted. We employed a plane-wave cutoff energy of 60 Ry and energy convergence criterion of $1.0\times10^{-14}$ Ry \AA$^{-1}$. We constructed maximally localized Wannier functions (MLWFs) from d orbitals of Nb atoms and p orbitals of Se atoms using Wannier90 code~\cite{Mostofi2014}. Lattice dynamics were computed in PHONOPY by density functional perturbation theory (DFPT)~\cite{Parlinski1997} on a coarse q-mesh ($12\times12\times4$), and the DFPT results were combined with the MLWFs to evaluate electron–phonon matrix elements in the EPW code~\cite{Giustino2007} on a very dense q-mesh with $32\times32\times16$. At each strain value, lattice vectors were scaled accordingly and internal coordinates were fully relaxed before anisotropic electron-phonon coupling calculation. All phonon spectra show no imaginary modes and remain robust under strain.

We have performed computational calculations in EPW to get insight into electron-phonon coupling strength~\cite{Sio2019} and superconductor critical temperature~\cite{Heil2017}. Based on the Anisotropic Migdal-Eliashberg theory, which is a generalization of BCS theory to include the retardation effect~\cite{Eliashberg1960}, such that the electron-phonon coupling strength has the following expression:

\begin{equation}
    \lambda(nk, mk', \omega_j - \omega_j') =
    \int_{0}^{\infty} \frac{2\Omega}{(\omega_j - \omega_j')^2 + \Omega^2} \alpha^2 F(nk, mk', \Omega) \, \mathrm{d}\Omega.
    \label{eq:lambda_expression}
\end{equation}
where $\alpha^2 F$ is the fully anisotropic Eliashberg spectral function, which is
included to capture multi-band effects. The superconductor transition temperature
is subjected to the ratio between the order parameter and the renormalization function, which is obtained by following sea t of coupled non-linear equations:

\begin{equation}
    Z_{nk}(i\omega_j) = 1 + \frac{T}{\omega_j} \sum_{mk'j'} \frac{\omega_j Z_{mk'}(i\omega_{j'})}{\theta_{mk'}(i\omega_{j'})}
    \times \left\{ \frac{\lambda(nk, mk', \omega_j - \omega_{j'})}{N(\varepsilon_F)} - V_{nk-mk'}(i\omega_j - i\omega_{j'}) \right\}
\end{equation} \begin{equation}
    \chi_{nk}(i\omega_j) = -T\sum_{mk'j'} \frac{\varepsilon_{mk'}-\varepsilon_F + \chi_{mk'}(i\omega_{j'})}{\theta_{mk'}(i\omega_{j'})}
    \times \left\{ \frac{\lambda(nk, mk', \omega_j-\omega_{j'})}{N(\varepsilon_F)} - V_{nk-mk'}(i\omega_j-i\omega_{j'}) \right\}
\end{equation} \begin{equation}
    \phi_{nk}(i\omega_j) = T\sum_{m'k'} \frac{\phi_{m'k'}(i\omega_j)}{\theta_{m'k'}(i\omega_j)}
    \times \left\{ \frac{\lambda(nk, m'k', \omega_j-\omega_j)}{N(\varepsilon_F)} - V_{nk-m'k'}(i\omega_j-i\omega_j) \right\}
\end{equation}

The Fermi energy of the system with electron number $N_e$ stems from:

\begin{equation}
    N_e = 1 - 2T \sum_{mk'_j} \frac{\varepsilon_{mk'} - \varepsilon_{\text{F}} + \chi_{mk'}(i\omega_j)}{\theta_{mk'}(i\omega_j)}.
    \label{eq:placeholder}
\end{equation}
Finally, the superconducting gap $\Delta_{nk}$ can be obtained by the following equation:
\begin{equation}
    \Delta_{nk}(i\omega_j) = \frac{\phi_{nk}(i\omega_j)}{Z_{nk}(i\omega_j)}
    \label{eq:1}
\end{equation}
The maximum temperature at which these coupled equations admit a non-trivial solution ($\Delta_{nk} \neq 0$) represents the superconducting critical temperature $T_c$.

\subsection*{TDGL simulation}
In order to reveal the influence of the substrate on the superconducting properties, we use the finite volume method to simulate the generalized time-dependent Ginzburg-Landau (TDGL) equation. %
The evolution of a superconductor can be described by the TDGL equation,
the order parameter $\psi$ evolves in dimensionless units as follows~\cite{Kramer1978, Jing2018, Bishop-VanHorn2023},
\begin{equation}
    \frac{u}{\sqrt{1 + \gamma^2|\psi|^2}}\left(\frac{\partial}{\partial t} + i\mu + \frac{\gamma^2 \partial|\psi|^2}{2 \partial t}\right)\psi = (\epsilon - |\psi|^2)\psi + (\nabla - i\mathbf{A})^2\psi.
\end{equation}

The quantity $(\nabla - i\mathbf{A})^2\psi$ is the covariant Laplacian of $\psi$, which is used in place of an ordinary Laplacian in order to maintain gauge-invariance.
Similarly, $(\frac{\partial}{\partial t} + i\mu)\psi$ is the covariant time derivative of $\psi$. The real-valued parameter $\epsilon(\mathbf{r}) = 1 - T/T_c(\mathbf{r}) \in [-1, 1]$ is the temperature dependent disorder parameter with temperature normalized to the spatially-varying critical temperature $T_c(\mathbf{r})$.
The electric potential $\mu(\mathbf{r},t)$ evolves according to the Poisson equation:
\begin{equation}
    \nabla^2\mu = \nabla \cdot \text{Im}[\psi^*(\nabla - i\mathbf{A})\psi] = \nabla \cdot \mathbf{J}_s,
\end{equation}
where $\mathbf{J}_s$ is the dissipationless supercurrent density. The total current density is the sum of the supercurrent density $\mathbf{J}_s$ and the normal current density $\mathbf{J}_n$:
\begin{equation}
    \mathbf{J} = \mathbf{J}_s + \mathbf{J}_n = \text{Im}[\psi^*(\nabla - i\mathbf{A})\psi] - \nabla\mu,
\end{equation}
where $(\nabla - i\mathbf{A})\psi$ and $\psi^*$ are the covariant gradient and the complex conjugate of $\psi$, respectively. Eq. 2 results from applying the current continuity equation $\nabla \cdot \mathbf{J} = 0$ to Eq. 3.
The total heat source can be obtained by the Helmholtz free energy and the heat created by normal current density\cite{Hernandez2008, Jing2018},
\begin{equation}
    W_{total} = 2\left(\frac{\partial \mathbf{A}}{\partial t}\right)^2 + \frac{2u}{\sqrt{1 + \gamma^2|\psi|^2}}\left[\left(\left|\frac{\partial \psi}{\partial t}\right|\right)^2 + \frac{\gamma^2}{4}\left(\frac{\partial |\psi|^2}{\partial t}\right)^2\right] + \boldsymbol{J}_n^2.
\end{equation}

The constant $u = \pi^4/14\zeta(3) \approx 5.79$ is the ratio of relaxation times for the amplitude and phase of the order parameter in very dirty superconductors ($\zeta(x)$ is the Riemann zeta function), and $\gamma=1$ parametrizes the strength of inelastic scattering as described above.
In the clean limit, $\gamma=0$ and $u=1.0$ are set.
The TDGL simulation is performed by using the py-tdgl package~\cite{Bishop-VanHorn2023}, with the modification of adding the thermal diffusion equation~\cite{Jing2018},
\begin{equation}
    C_{eff} \frac{\partial T}{\partial t} = \nabla \cdot (k_{eff} \nabla T) + \frac{1}{2}W_{total} - \eta (T-T_0),
\end{equation}
where, $C_{eff}=1.0$ is the effective heat capacity, $K_{eff}=0.02$ is the effective thermal conductivity, which is in the same order as that reported in Ref.~\cite{Jing2018}, $\eta$ is the heat transfer coefficient of the substrate.


\section*{Acknowledgements}
The authors express gratitude for the facilities and technical support provided by Aalto University at OtaNano-Nano microscopy Center (OtaNano/NMC), OtaNano/LTL, and Micronova cleanroom. We thank Prof. Tero Heikkilä for insightful theoretical discussions on the DFT and TDGL simulations. This work was supported by the Research Council of Finland (352780, 352926,352930, 353364, 360411, 359009, 365686,and 367808), the Research Council of Finland Flagship Programme (320167, PREIN), Finnish Quantum Flagship project (358877, Aalto), the EU H2020-MSCA-RISE-872049 (IPN-Bio), R. Fang acknoledges the European Union's Horizon research and innovation program (101155102,Q-LAMP).

\section*{Author contributions}
Z. Sun and H. Fang conceived the idea. R. Fang, P. Hakonen and T. Elalaily conceived the design and analysis of experiment. R. Fang and H. Fang fabricated the devices with the help of  S. Xiao, J. Park, H. Choi and F. Ali. R. Fang, Y. Huang, Y. Chaudry and Y. Zhou performed the experiments. C. Yu and J. Lado performed the DFT calculations. R. Fang performed the TDGL simulations. All authors contributed to the writing of the article.

\section*{Data availability}
The data that support the findings of this study are available from the corresponding authors upon reasonable request. 

\bibliography{Collection}

\end{document}


\title{Supplementary for Controllable Superconductivity in Suspended van der Waals Materials}


\author[1]{\fnm{Ruihuan} \sur{Fang}}
\author[2]{\fnm{Cuiju} \sur{Yu}}
\author[1]{\fnm{Youqiang} \sur{Huang}}
\author[2]{\fnm{Tosson} \sur{Elalaily}}
\author[2]{\fnm{Yuvraj} \sur{Chaudhry}}
\author[1]{\fnm{Yaoqiang} \sur{Zhou}}
\author[4]{\fnm{Andres} \sur{Castellanos-Gomez}}
\author[3]{\fnm{Sanshui} \sur{Xiao}}
\author[5]{\fnm{Jiwon} \sur{Park}}
\author[5]{\fnm{Hyunyong} \sur{Choi}}
\author[1]{\fnm{Fida} \sur{Ali}}
\author*[3]{\fnm{Hanlin} \sur{Fang}}\email{hanfang@dtu.dk}
\author*[2]{\fnm{Jose} \sur{Lado}}\email{jose.lado@aalto.fi}
\author*[2]{\fnm{Pertti} \sur{Hakonen}}\email{pertti.hakonen@aalto.fi}
\author*[1]{\fnm{Zhipei} \sur{Sun}}\email{zhipei.sun@aalto.fi}

\affil[1]{\orgdiv{QTF Centre of Execellence, Department of Electronics and Nanoengineering}, \orgname{Aalto University}, \orgaddress{\city{Espoo}, \postcode{02150}, \country{Finland}}}

\affil[2]{\orgdiv{Department of Applied Physics}, \orgname{Aalto University}, \orgaddress{\city{Espoo}, \postcode{02150}, \country{Finland}}}

\affil[3]{\orgdiv{Department of Electrical and Photonics Engineering}, \orgname{Technical University of Denmark}, \orgaddress{\city{Copenhagen}, \postcode{610101},\country{Denmark}}}

\affil[4]{\orgdiv{2D Foundry Research Group}, \orgname{Instituto de Ciencia de Materiales de Madrid (ICMM-CSIC)}, \orgaddress{\postcode{28049}, \city{Madrid}, \country{Spain}}}

\affil[5]{\orgdiv{Department of Physics}, \orgname{Seoul National University}, \orgaddress{\city{Seoul}, \postcode{08826}, \country{Korea}}}

\maketitle

\newpage

\section{Micrograph of NbSe$_2$ devices}

\begin{figure}[!h]
    \centering
    \includegraphics[width=0.8\linewidth]{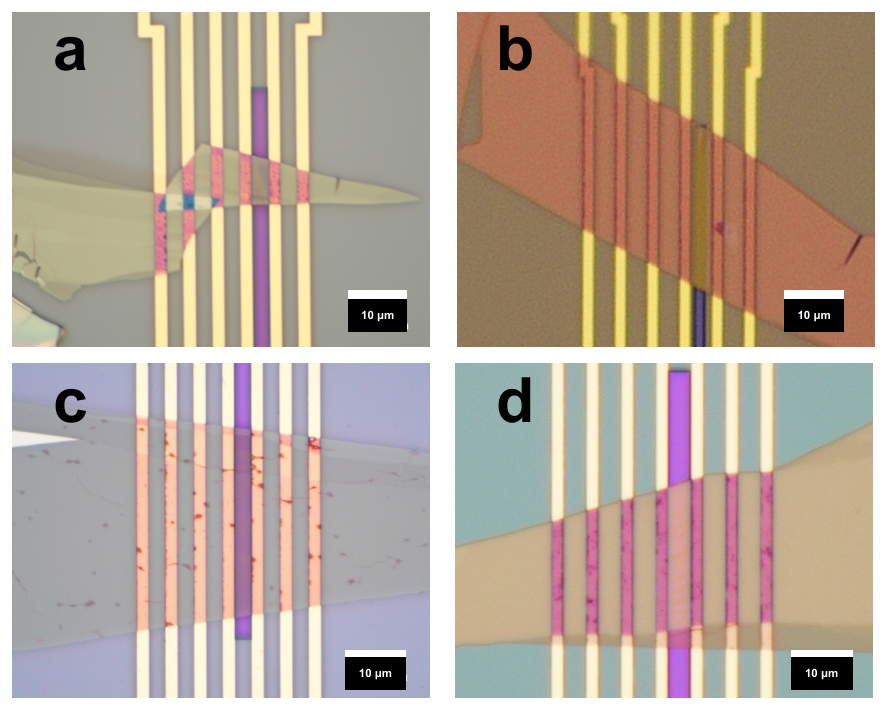}
    \caption{\textbf{Optical micrographs of suspended NbSe$_2$ devices.}
    (a) Device~1 with an electrode width of $2~\upmu\mathrm{m}$ (scale bar: $10~\upmu\mathrm{m}$), inter-electrode gap of $3~\upmu\mathrm{m}$, suspended width of $w_s\approx 3~\upmu\mathrm{m}$, and NbSe$_2$ thickness of $\sim 13~\mathrm{nm}$.
    (b) Device~2 with electrode width of $2~\upmu\mathrm{m}$ (scale bar: $10~\upmu\mathrm{m}$), inter-electrode gap of $2~\upmu\mathrm{m}$, suspended width of $w_s\approx 2~\upmu\mathrm{m}$, and NbSe$_2$ thickness of $10~\mathrm{nm}$.
    (c) Device~3 with electrode width of $~2~\upmu\mathrm{m}$ (scale bar: $10~\upmu\mathrm{m}$), inter-electrode gap of $3~\upmu\mathrm{m}$, suspended width of $w_s\approx 3~\upmu\mathrm{m}$, and NbSe$_2$ thickness of $5~\mathrm{nm}$.
    (d) Device~4 with electrode width of $2~\upmu\mathrm{m}$ (scale bar: $10~\upmu\mathrm{m}$), inter-electrode gap of $4~\upmu\mathrm{m}$, suspended-hole width of $w_s\approx 4~\upmu\mathrm{m}$, and NbSe$_2$ thickness of $15~\mathrm{nm}$.}
    \label{fig:schematic}
\end{figure}

Figure~\ref{fig:schematic} presents optical micrograph of four suspended NbSe$_2$ devices studied in the main text. The detailed geometric parameters are summarised in Table~\ref{tab:device_parameters}.

\begin{table}[!h]
  \centering
  \caption{Geometric parameters of the four suspended NbSe$_2$ devices.
  All lateral dimensions are in $\upmu$m and thickness in nm.}
  \label{tab:device_parameters}
  \begin{tabular}{lcccc}
    \toprule
    \textbf{Device No.} & \textbf{Suspended width $w_s$} & \textbf{Flake width $w_f$} & \textbf{Thickness} & \textbf{Figure ref.} \\
     & \textbf{($\upmu$m)} & \textbf{($\upmu$m)} & \textbf{(nm)} & \\
    \midrule
    1 & $\sim$3   & $\sim 8$  & $\sim 13$ & Fig.~\ref{fig:schematic} a \\
    2 & $\sim$2   & $\sim 25$ & $\sim 11$ & Fig.~\ref{fig:schematic} b \\
    3 & $\sim$3   & $\sim 30$ & $\sim 5$  & Fig.~\ref{fig:schematic} c \\
    4 & $\sim$4   & $\sim 25$ & $\sim 11$ & Fig.~\ref{fig:schematic} d \\
    \bottomrule
  \end{tabular}
\end{table}

\section{Strain in collapsed device}
\begin{figure}[!h]
    \centering
    \includegraphics[width=1.0\linewidth]{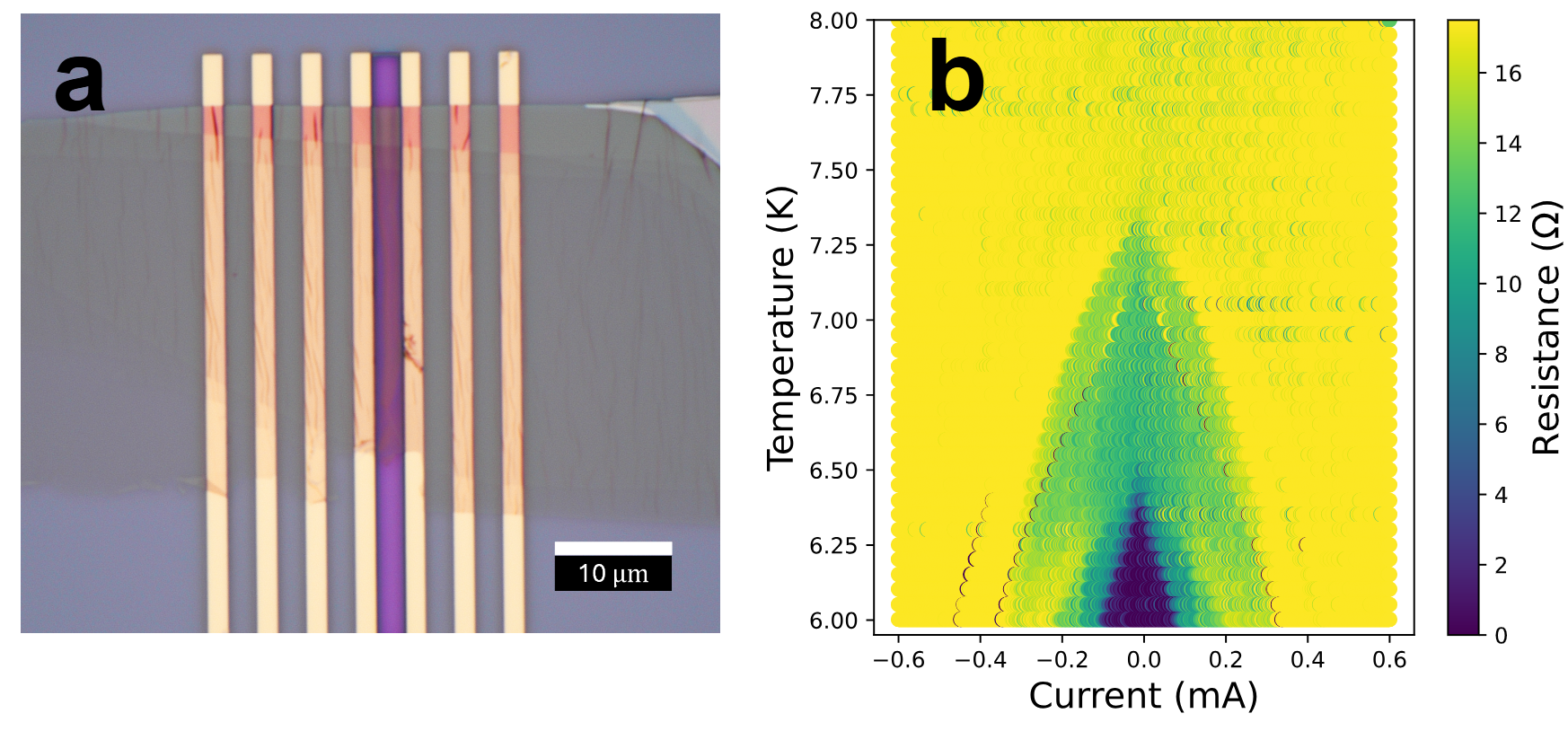}
    \caption{\textbf{Characterization of NbSe$_2$ devices under strain.}
    (a) Optical micrograph of a suspended NbSe$_2$ device with electrode width of $2~\upmu\mathrm{m}$ (scale bar: $10~\upmu\mathrm{m}$), inter-electrode gap of $3~\upmu\mathrm{m}$, suspended width of $w_s=3~\upmu\mathrm{m}$, and NbSe$_2$ thickness of $\sim 3~\mathrm{nm}-6~\mathrm{nm}$.
    (b) Differential resistance as a function of current and temperature.}
    \label{fig:collapse}
\end{figure}

The micrograph of the device that collapsed into an etched trench is shown in Fig.~\ref{fig:collapse}a, and the resistance as a function of bias current and temperature is shown in Fig.~\ref{fig:collapse}b. The collapsed flake has a thickness of \(\sim 3\)--\(6\,\mathrm{nm}\). Consequently, the suspended region can host segments with different local thickness and strain, leading to variations in the critical temperature \(T_\mathrm{c}\).

In principle, the multiple \(T_\mathrm{c}\) values extracted by sigmoid fits in Fig.~2b of main text could arise from two scenarios:
(i) under a large injected current, parallel segments with different \(T_\mathrm{c}\) switch to the normal state at different temperatures; or
(ii) serial weak links with different local \(T_\mathrm{c}\) produced by nonuniform strain (and thickness).

From Fig.~\ref{fig:collapse}b we identify three distinct current-driven transitions that occur at a characteristic current scale of \(\sim 0.1\,\mathrm{mA}\), which is far larger than the excitation current used for the temperature-dependent resistance measurement in Fig.~2b (\(1\,\upmu\mathrm{A}\)). Therefore, none of the segments are forced normal by the measurement current during \(R(T)\); current-induced sequential switching (scenario~i) can be ruled out for Fig.~2b. We thus attribute the observed multi-step transition temperatures primarily to spatial inhomogeneity—especially strain—across the collapsed flake (scenario~ii).

\section{Strain estimation using membrane's sagging }
\begin{figure}[!h]
    \centering
    \includegraphics[width=1.0\linewidth]{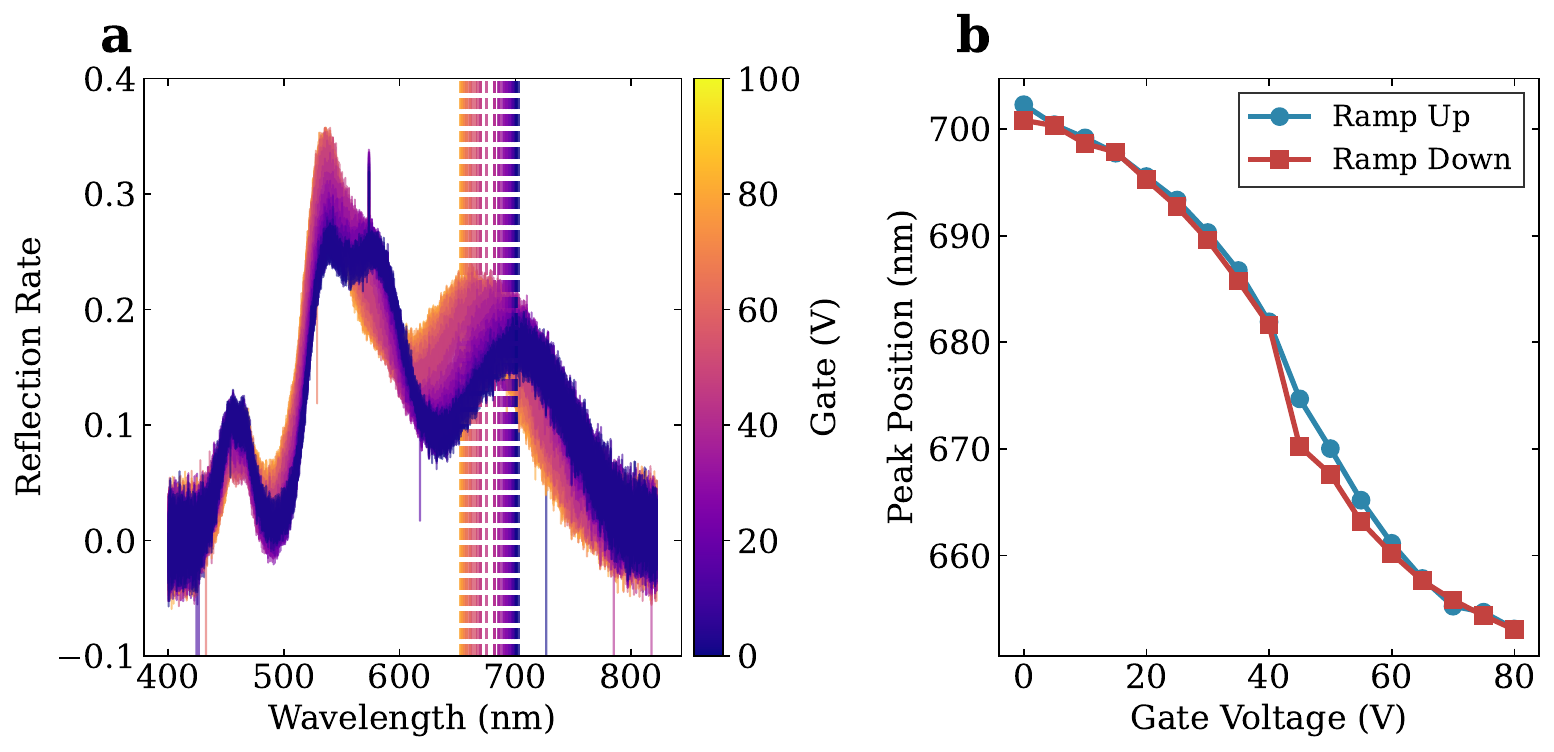}
    \caption{\textbf{Gate-dependent spectral peak shift.}
    (a) Reflection spectra measured at various gate biases; dashed lines indicate the peak wavelengths.
    (b) Peak wavelength versus gate voltage, showing a systematic shift.
    Device geometry is shown in Fig.~\ref{fig:schematic}a; the suspended width is \(w_s=3~\mu\mathrm{m}\).
    }
    \label{fig:spectrum}
\end{figure}

We model the suspended region as a rectangular membrane bridge with two opposite edges clamped (separated by distance \(L\)) and the other two edges free. For small deflection and membrane-dominated behavior, the first-mode transverse deflection along the clamped span is
\begin{equation}
w(x) = w_0 \sin\!\left(\frac{\pi x}{L}\right),
\end{equation}
where \(w_0\) is the center sag.

The corresponding average axial strain along the clamped span is approximated by
\begin{equation}
\varepsilon_{\mathrm{avg}} \approx \frac{\pi^2}{4}\left(\frac{w_0}{L}\right)^2 .
\end{equation}

For the geometry in Fig.~\ref{fig:schematic}a and corresponding to the gate tunability critical temperature shown in Fig.~2c and d, the clamped span \(L = 3~\mu\mathrm{m}\) (equal to the suspended width \(w_s\)).
The reflection spectrum with gate voltage is shown in Fig.~\ref{fig:spectrum}a. As Fig.~\ref{fig:spectrum}b shows, the peak shift $\sim50$ nm, means that center sag \(w_0 \sim 50~\mathrm{nm} = 0.05~\mu\mathrm{m}\). Therefore we can calculate the strain with

\begin{equation}
\begin{aligned}
\frac{w_0}{L} &= \frac{0.05}{3} \approx 1.67\times 10^{-2},\\
\varepsilon_{\mathrm{avg}} &\approx \frac{\pi^2}{4}\left(1.67\times 10^{-2}\right)^2
\approx 6.85\times 10^{-4} = 0.0685\%.
\end{aligned}
\end{equation}


\section{Leakage current influence}
\begin{figure}[!h]
    \centering
    \includegraphics[width=1.0\linewidth]{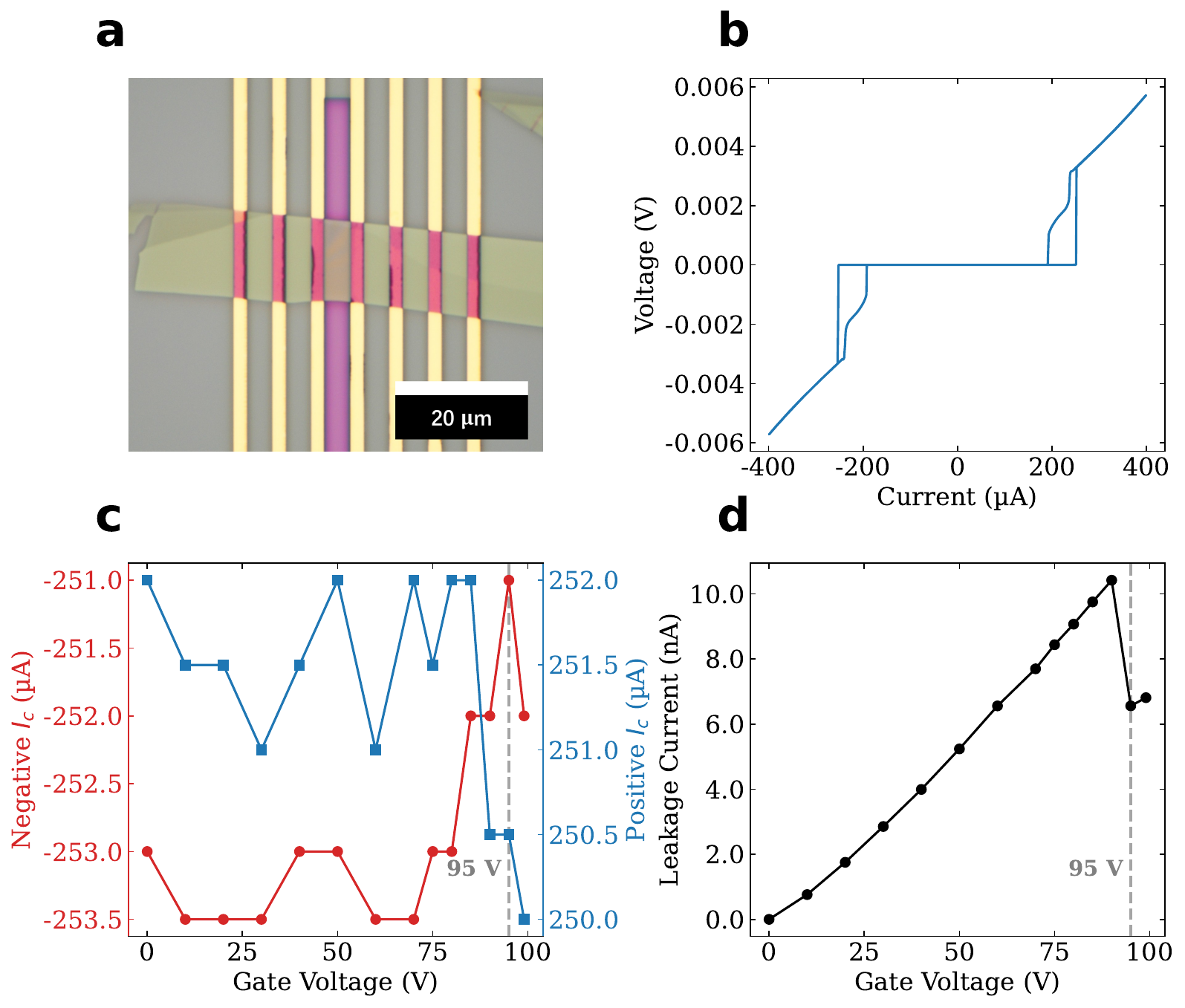}
    \caption{\textbf{Gate–device leakage characterization.}
    (a) Optical micrograph of the suspended device.
    (b) Representative two-terminal $V$–$I$ trace of the channel.
    (c) Positive/negative switching currents versus gate voltage $V_g$ in temperature of 15 mK.
    (d) Gate–device leakage current versus $V_g$, showing a monotonic increase and drop down when 95 V.}
    \label{fig:leakage}
\end{figure}

Applying a gate voltage can deform the suspended NbSe$_2$ flake; in parallel, any gate-leakage current produces Joule heating that could, in principle, modify superconducting parameters such as the critical temperature \(T_\mathrm{c}\) and critical current \(I_\mathrm{c}\). To examine these effects, we studied the device shown in Fig.~\ref{fig:leakage}a (thickness \(\sim 13\,\mathrm{nm}\)); its \(I\)--\(V\) characteristics are presented in Fig.~\ref{fig:leakage}b. The gate-leakage current measured as a function of gate voltage on the same device (Fig.~\ref{fig:leakage}d) reaches a maximum of \(\sim 10\,\mathrm{nA}\) and then drops near \(V_\mathrm{g}\approx 95\,\mathrm{V}\), which we attribute to a slip event that alters the leakage path. Notably, although the leakage current decreases at this point, the absolute value of \(I_\mathrm{c}\) remains suppressed. This behavior indicates that the observed reduction of superconducting strength is not dominated by leakage-induced heating; instead, the gate dependence of \(I_\mathrm{c}\) is more consistent with electrostatic and/or strain-induced modulation of the superconducting state.